\documentstyle[12pt,epsf,epsfig]{article}
\def\beq{\begin{equation}}
\def\eeq{\end{equation}}
\def\bea{\begin{eqnarray}}
\def\eea{\end{eqnarray}}
\def\ba{\begin{array}}
\def\ea{\end{array}}
\def\gappeq{\mathrel{\rlap {\raise.5ex\hbox{$>$}}
{\lower.5ex\hbox{$\sim$}}}}
\def\permil{$\%\raise.20ex\hbox{$_0$}}
\def\lappeq{\mathrel{\rlap{\raise.5ex\hbox{$<$}}
{\lower.5ex\hbox{$\sim$}}}}
\begin{document}
\topmargin -1.0cm
\oddsidemargin -0.8cm
\evensidemargin -0.8cm
\pagestyle{empty}
\begin{flushright}
CERN-TH/96-192\\
hep-ph/9607383
\end{flushright}
\vspace*{5mm}
\begin{center}
{\Large\bf Anomalous U(1) as a mediator}\\
\vspace{0.5cm}
{\Large\bf  of Supersymmetry Breaking}\\
\vspace{2cm}
{\large\bf Gia Dvali and Alex Pomarol}\\
\vspace{.4cm}
{Theory Division, CERN}\\
{CH-1211 Geneva 23, Switzerland}\\
\end{center}
\vspace{2cm}
\begin{abstract}
We point out that an anomalous gauge U(1) symmetry is a natural
candidate for being the  mediator and messenger  of supersymmetry breaking.
It facilitates dynamical supersymmetry breaking even in the flat limit.
Soft masses are induced by both gravity and the U(1)
gauge interactions giving  an unusual  mass hierarchy in the
sparticle spectrum which  suppresses flavor violations.
This scenario does not suffer from the Polonyi problem.
\end{abstract}

\vfill
\begin{flushleft}
CERN-TH/96-192\\
July 1996
\end{flushleft}
\eject
\pagestyle{empty}
\setcounter{page}{1}
\setcounter{footnote}{0}
\pagestyle{plain}


\section{Introduction}
The origin of supersymmetry breaking remains an open question.
More important, for phenomenological purposes, it is to know how the 
breaking of supersymmetry is transmitted to the ordinary particles.
The most popular scenario arises in the context of supergravity.
In these theories supersymmetry is assumed to be broken 
in some isolated hidden sector and  transmitted to
the observable sector by  gravity \cite{nilles}. 
These models, however,  suffer from certain drawbacks. 
The
degeneracy of the scalar quarks needed to avoid large
 flavor changing neutral currents (FCNC) is not usually guaranteed at
low energies.
Also the breaking of 
supersymmetry results  in the non-flat limit 
leading to cosmological  disasters
 (the Polonyi problem \cite{poloni}).

In this letter we will consider 
an alternative scenario.
It is well known that  extra  U(1) factors 
 normally appear in effective field
theories arising from strings.
One  of these U(1) is usually anomalous. 
The cancellation of its anomalies occurs
by the Green-Schwarz  mechanism \cite{gs} and requires that
both hidden and observable fields transform non-trivially under 
this U(1). Thus,  this anomalous U(1) 
seems to be a natural new candidate for
transmitting the supersymmetry breaking from the hidden
to the observable sector. Here we will study this possibility.

Since the U(1) is anomalous, ${\rm Tr}{\bf Q} \not= 0$,
a Fayet-Iliopoulos term of ${\cal O}(M^2_P)$ is 
always generated \cite{witten}. This term 
facilitates the
breaking of supersymmetry in the flat limit, avoiding
the Polonyi problem. The scale of 
supersymmetry breaking
can be smaller than $M_P$ and can originate dynamically.
In the presence of 
gravity, 
realistic scalar and 
gaugino masses are induced 
in the observable sector.
We find that
the $D$-term contribution can be  larger than the
gravity mediated $F$-term contribution, resulting in a 
hierarchy of soft masses.
This is a
crucial difference with  the  conventional 
 hidden sector scenarios in supergravity models.
As we will show,  our model can  lead to a certain degree of
 squark degeneracy and  suppressed FCNC.
It also allows for an explanation of
the observed quark mass hierarchy 
($m_{t,b}\gg m_{u,d},m_{c,s}$) 
and predicts an inverse hierarchy for the squarks
  ($m^2_{\tilde u,\tilde d}
\simeq m^2_{\tilde c,\tilde s}\gg m^2_{\tilde t,\tilde b}$).

Anomalous U(1) have been considered before 
to predict the weak mixing angle \cite{ibanez},
 fermion \cite{u1} or sfermion \cite{scalar}
masses; in these  previous analysis, however,  
the anomalous U(1) does not play any role in the breaking
of supersymmetry. 

\section{Supersymmetry Breaking with  an Anomalous U(1)} 

Let us consider 
a pair of chiral superfields
$\phi^-$ and $\phi^+$ with charges equal to $ -1$ and $+ 1$ respectively
under a gauge  U(1).
We will assume that
there are other positively charged fields $Q_i$
 such that ${\rm Tr}{\bf Q} > 0$
and the U(1) is anomalous.
This results into the appearance
of a Fayet-Iliopoulos term  $\xi={\cal O}(M^2_P{\rm Tr}{\bf Q})$
\cite{witten}.
In string theories the generated Fayet-Iliopoulos term 
can be calculated and is given by \cite{fayetil}
\begin{equation}
 \xi = {g^2{\rm Tr}{\bf Q} \over 192\pi^2}M^2_P\, .
\end{equation}
The
$D$-term contribution to the effective potential takes the form
\begin{equation}
 {g^2 \over 2}D^2 = {g^2 \over 2}\left (\sum_i
q_i|Q_i|^2+|\phi^+|^2 - |\phi^-|^2+\xi\right)^2\, ,
\label{dtermcon}
\end{equation}
where $q_i$ is the U(1)-charge of the field $Q_i$.
If eq.~(\ref{dtermcon}) is the only term in the potential,
the vacuum expectation value (VEV) of $\phi^-$  adjusts
 to compensate $\xi$, and supersymmetry will not be broken.
 However, according to the old observation
by Fayet \cite{fayet}, 
this can lead to the spontaneous breakdown of the
supersymmetry if the
$\phi^-$ field has a non-zero mass term in the superpotential:
\begin{equation}
W = m\phi^+\phi^-\, .
\end{equation}
We will show below that such a mass term can
in fact be generated dynamically.
For the  moment, let us consider it 
 as a new input of the theory and look for its consequences.
Minimization of the potential 
shows that the VEVs of the scalar components are 
\begin{equation}
\langle\phi^+\rangle = 0,~~~\langle\phi^-\rangle^2 
= \xi- {m^2 \over g^2}\, ,
\label{vacuums}
\end{equation}
and the VEVs of the $F$- and $D$-components are given by
\begin{equation}
\langle F_{\phi^+}\rangle = m\sqrt{\xi- {m^2 \over g^2}},~~~
\langle F_{\phi^-}\rangle = 0,~~~ \langle D\rangle =\frac{m^2}{g^2}\, .
\label{vacuum}
\end{equation}
The spectrum of the theory is the following: (1) The Goldstone boson
$Im\phi^-$ is eaten up by the gauge field that gets a mass
$g\sqrt{\xi - {m^2 \over g^2}}$ \cite{argu}; 
(2) its superpartner $Re\phi^-$
gets a mass $g\sqrt{\xi -{m^2 \over g^2}}$ from the $D$-term and
becomes a member of the massive gauge superfield; 
(3) the complex scalar $\phi^+$ gets a squared-mass $2m^2$;
(4) one linear combination
of the chiral fermions and 
the gaugino  gets a Dirac mass $g\sqrt{\xi- {m^2 \over 2g^2}}$, 
whereas the  orthogonal
combination is the massless Goldstino.

Let us  now embed this model  in a supergravity theory.
It is easy to show that the
broken global supersymmetry cannot be restored  by the supergravity
interactions. 
This is  because
an unbroken
supergravity with vanishing vacuum energy
implies $\langle W\rangle = 0$ and therefore 
 that all $\partial_\phi W$ and $D_A$ vanish too; this
contradicts the initial assumption that supersymmetry was broken
in the flat limit.  
Under supergravity,  
the VEVs of the fields will be shifted from eqs.~(\ref{vacuums})
and (\ref{vacuum}),
but the relation
\begin{equation}
{\langle F^2 \rangle \over \langle D \rangle} \sim\xi\, ,
\label{relation}
\end{equation}
will still hold.

\section{The Sparticle Spectrum}

In a  supergravity theory the  
 supersymmetry
breaking is
communicated by gravity from the hidden sector ($\phi^+,\phi^-$)
to the observable 
sector ($Q_i$).
The scalar masses receive  contributions  of order
\begin{equation}
m^2_Q\simeq \frac{\langle F_{\phi^+}\rangle^2}{M^2_P}
\simeq\frac{m^2\xi}{M^2_P}\simeq \varepsilon m^2\, ,
\label{fterm}
\end{equation}
where $\varepsilon\equiv \xi/M^2_P$ that in string theories takes the value
$\varepsilon=g^2{\rm Tr}{\bf Q}/192\pi^2$.
These contributions are, 
in principle,  non-universal, since they depend on the
K\"ahler potential \cite{nilles}.
The gaugino masses can arise from the operator
\begin{equation}
\int d^2\theta \frac{\phi^+\phi^-}{M^2_P}W_aW_a\, ,
\label{operator}
\end{equation}
where $W_a$ is the superfield that contains the gauge field strength
of the standard model SU$(a)$ group, $a=1,2,3$.
Thus, gaugino masses  are given by
\begin{equation}
m_\lambda\simeq\frac{\langle F_{\phi^+}\phi^-\rangle}{M^2_P}\simeq
\varepsilon m\, .
\label{gaugi}
\end{equation}
Notice that the presence of the field  $\phi^-$  with a VEV of order $M_P$
is crucial to give acceptable gaugino masses from 
the operator eq.~(\ref{operator}).
The absence of this field in other models in which
supersymmetry is also broken in the 
flat limit, leads to very light gauginos \cite{ads} (see
however ref.~\cite{nelson}). 
In string theories the operator eq.~(\ref{operator}) can only be
 induced at the one-loop  level since only the dilaton couples to $W_aW_a$
at the tree level.  Larger contributions to the gaugino masses, however, 
can arise from integrating out heavy states as we will
show in the next section.

Since in our scenario $\langle D\rangle$  is different from zero,
extra contributions to the scalar masses arise from
the $D$-term  for  fields
that transform under the anomalous U(1).
{}From eqs.~(\ref{dtermcon}) and (\ref{vacuum}), these  are given by
\begin{equation}
\Delta m^2_{Q_i}=q_i\, m^2\, .
\label{dterm}
\end{equation}
Notice that these contributions can be much larger than the $F$-term
contributions eq.~(\ref{fterm}) if $\varepsilon\ll 1$. Thus, 
this scenario allows for 
a hierarchy of soft masses:
\begin{equation}
\Delta m^2_Q>m^2_Q>m^2_\lambda\, .
\end{equation} 
This is different from models 
in which the U(1) does not play any role in the breaking of supersymmetry.
In those models the $D$-term contribution to
the scalar masses is always of the same order
as  the $F$-term contribution \cite{scalar}.
The spectrum eqs.~(\ref{fterm}), (\ref{gaugi}) and
(\ref{dterm}) is a general feature
of this {\it hybrid} scenario where the breaking of supersymmetry 
is transmitted by both gravity and  U(1)-gauge interactions and is due to
 the  generic relation eq.~(\ref{relation}).
This  allows for a solution to the 
supersymmetric flavor problem, {\it i.e.}
the required degeneracy  between  the first and 
second family squarks $\delta m^2_{Q}/m^2_Q\ll 1$. 
If  these two families of squarks transform non-trivially 
under the U(1), they receive the universal contribution of eq.~(\ref{dterm})
which, for  $\varepsilon\ll 1$, can be much larger
than the non-universal  contribution eq.~(\ref{fterm}) and 
therefore
\begin{equation}
\frac{\delta m^2_{Q}}{m^2_Q}\simeq{\varepsilon}\ll 1\, .
\end{equation} 
Decreasing $\varepsilon$ increases not only the degeneracy of
the first two family squarks,
but also increases their soft masses with respect to the other ones and then
further suppresses the supersymmetric FCNC contributions.
Obviously, $\varepsilon$ cannot be much smaller than 1, otherwise
the gaugino masses obtained 
from (\ref{gaugi}) are too small.
The best scenario that we envisage is to have the three 
quark families transforming
under the U(1) as $\{1,1,0\}$ respectively \cite{pos}.
{}For reasonable values of 
$\varepsilon=g^2{\rm Tr}{\bf Q}/192\pi^2\simeq 10^{-2}$, we get 
for $m\simeq 5$ TeV:
\begin{equation}
m_\lambda\simeq 50\ {\rm GeV}\ ,\  
m_{Q_3}\simeq 500\ {\rm GeV}\ ,\  
m_{Q_{1,2}}\simeq 5\ {\rm TeV}\, .\label{splitting}
\end{equation}
This is a spectrum very similar to that in ref.~\cite{pt}. The
 FCNC are  suppressed enough. Furthermore,
this scenario provides a solution to 
the  supersymmetric CP problem
\cite{edm}. 
This is  because the first
 family of squarks are so  heavy that
 their contribution to the electric dipole moment
of the neutron is small,
even if 
the CP-violating phases are of ${\cal O}(1)$.
It is important to remark that
the large mass splitting eq.~(\ref{splitting})
 does not lead to a naturalness 
problem, since the first two families are almost decoupled from the Higgs
\cite{dg,pt}.

The above anomalous U(1) could also play a role in explaining 
the fermion masses in the 
same spirit as in ref.~\cite{u1}. Here, however,
we are constrained to have the 
first two families with equal U(1) charges
(in order to avoid too large FCNC) \cite{pos}.
Although a complete model will not be attempted in this letter,
it is interesting to note that
if,  as we mentioned above, the Higgs and the $3^{rd}$ family are
neutral under this U(1) but the $1^{st}$ and $2^{nd}$
ones are charged, a tree-level mass
is only allowed  for the $3^{rd}$ family, explaining why the top and
bottom masses are much larger than the others. This scenario 
relates the mass hierarchy of the quarks to that 
in eq.~(\ref{splitting}) for the squarks.

It is worth to point out that, contrary to most of the flavor models,
our scenario allows for gauging extra flavor symmetries,
since the universal contribution eq.~(\ref{dterm}) 
dominates over any other
non-universal $D$-term contribution.

\section{A Scenario of  Dynamical Supersymmetry Breaking}

Up to now we have assumed that
$m\sim 1$ TeV  is just a  new scale in the model.
In this section we will show
that this scale can be generated dynamically.
We only need a   gauge group that at some intermediate scale
$\Lambda$ becomes  strongly interacting and  leads to  a field
condensation.

The simplest example is an SU(2) group with two doublets $\Phi$ and 
$\bar\Phi$, neutral under the anomalous U(1). 
At energies below the scale $\Lambda$, the low-energy effective 
theory can be described in terms of the gauge-invariant quantity
$X\equiv \Phi\bar\Phi$ \cite{ads}. 
The superpotential is given by
\begin{equation}
W=\lambda\frac{X}{M_P}\phi^+\phi^-+\frac{\Lambda^5}{X}\, ,
\end{equation}
where the first  term  has been assumed to be present
in the classical theory; the second term is generated 
non-perturbatively by instantons \cite{ads}. 
If no Fayet-Iliopoulos term is present in the theory,
the vacuum has a run-away behaviour, 
 $X\rightarrow\infty$ with $\phi^+,\phi^-\rightarrow 0$.
However, when the  U(1) $D$-term of eq.~(\ref{dtermcon}) is considered,
the field $\phi^-$ is forced to get a VEV and drives $X$ to a 
value around $\Lambda$. This generates the effective scale 
$m=\lambda\langle X\rangle/M_P$ and the 
breaking of supersymmetry. 
The only difference with respect to the model
of sect.~2
is that  $\phi^+$ now gets a VEV of order $\sqrt{\xi}$ 
and then $\langle F_{\phi^-}\rangle\sim  m\sqrt{\xi}$.
A new contribution to the gaugino masses  can now arises
from the operator
\begin{equation}
\frac{1}{16\pi^2}\int d^2\theta 
\frac{\phi^-}{\sqrt{\xi}}W_aW_a\, ,
\label{axion}
\end{equation}
that can be induced if extra heavy matter fields 
(transforming under the
standard model group)  are present and get their  masses from   couplings
to $\phi^-$. 
It can be shown that these couplings do not modify
the supersymmetry-broken vacuum. 
Although the operator eq.~(\ref{axion}) is
suppressed by a  one-loop factor, it is enhanced with respect to
the  gravity-induced operators since $\sqrt{\xi}< M_P$.
Eq.~(\ref{axion}) generates a mass term for  the gauginos
given by
\begin{equation}
m_\lambda\simeq\frac{1}{16\pi^2}\frac{\langle F_{\phi^-}\rangle}
{\sqrt{\xi}}
\simeq \frac{m}{16\pi^2}\, ,
\end{equation}
that can be as large as  eq.~(\ref{gaugi}).

The simplicity of this dynamical model resides in the fact that 
the strongly interacting gauge group  is only needed for
generating the small scale $m$ 
and not for  breaking the supersymmetry by itself
as in ref.~\cite{ads}.
Here it is the Fayet-Iliopoulos term 
that plays the new  and crucial role of triggering 
the breaking of supersymmetry.

\section{The  Polonyi Problem}

Perhaps the main cosmological difficulty of the
supergravity models with a conventional  hidden sector
is the Polonyi problem \cite{poloni}. This arises
because  models in which supersymmetry gets restored in the flat limit
predict  light ${\cal O}(m_{3/2})$ scalar particles with
VEVs of  ${\cal O}(M_P)$, with an extremely flat potential and  $1/M_P$
suppressed interactions. 
In the early universe these
fields are expected to sit far away from their
present (zero-energy) vacua.  The reason is that in the early universe
(during inflation or in the heat bath) these flat directions
get  large soft masses equal to $\alpha H^2$, 
where $H$ is  the Hubble parameter and
 $\alpha$ is a number of order 1 that
depends on the details of the cosmological scenario \cite{hub}.
For  particles with non-zero VEVs this leads, almost for
sure,  to a  classical displacement from the present vacuum at
the early times ($\Delta \sim M_P$) and to the subsequent coherent
oscillations around the true minimum after  inflation. The amplitude
and consequently the energy stored in the oscillations is determined by
the initial deviation and will overclose the universe if the
displacement is larger than $\sim 10^{-9}M_P$ \cite{poloni}.
For $\alpha > 0$ the displacement
is generically given by the value of the present VEV, whereas for
$\alpha < 0$ it can be much larger. Therefore, a light decoupled
scalar with a VEV larger than $10^{-9}M_P$ is problematic, whereas
scalars with smaller VEVs (at present) can be diluted by  inflation.
Now it is clear why the Polonyi problem can be overcome in  theories
with flat space supersymmetry breaking. Such theories
do not necessarily require scalars with large VEVs and vanishing mass
in the globally supersymmetric limit. In our models, the
field that gets a VEV of order $M_P$ is heavy; it is eaten up 
by the massive U(1)-gauge superfield.

\section{Conclusions}

\noindent$\bullet$ 
We pointed out that an anomalous gauge U(1) symmetry is a natural 
candidate for being the  mediator and messenger  of supersymmetry breaking.
It allows for  simple models of
dynamical supersymmetry breaking  in the flat limit.

\noindent$\bullet$  These models can be embedded in a supergravity 
theory and generate realistic scalar and gaugino soft masses.
The supersymmetry breaking is communicated by gravity
and the gauge U(1). This {\it hybrid} scenario allows for  
a solution to the supersymmetric flavor  and  CP problem.
The resulting phenomenology 
is very different from that of the usual models with universal
soft masses \cite{pt}.

\noindent$\bullet$ Since supersymmetry is broken in the flat limit,
there is no  Polonyi problem. All the hidden sector fields
are either very massive or get VEV below the Planck scale.

\vspace{.6cm}

It is a pleasure to thank  Gian Giudice, Amit Giveon,
Luis Ib\'a\~nez, Fernando Quevedo and Misha Shifman
 for very useful discussions.

{\it Note added}: After submitting this paper, we learned about a related
 work by P. Bin\'etruy and E. Dudas, preprint hep-th 9607172. 
We thank E. Dudas for comments.


\begin{thebibliography}{99}
\bibitem{argu} In string theories the eaten up Goldstone is a superposition
of $Im\phi^-$ and the model-independent axion (partner of the
dilaton) \cite{fayetil}.
\bibitem{pos} One can consider other possibilities such as 
equally charging only the left-handed squarks of the first two families.
This can be enough to avoid the FCNC constraints coming from the real
part of the $K^0$--$\bar K^0$ mass mixing if the corresponding right-handed
 squarks are heavier than  $\sim$ 5 TeV.
\end{thebibliography}
\end{document}